\documentclass[letter,twocolumn]{jpsj2} %% for letters
\usepackage{amsmath}
\usepackage{bm}
\usepackage{lscape}

\title{A Novel Ordered Phase in SrCu$_{2}$(BO$_{3}$)$_{2}$ under High Pressure}

\author{Takeshi WAKI$^{1}$\thanks{E-mail address: twac@issp.u-tokyo.ac.jp},
Koichi ARAI$^{1}$\thanks{Present address: Hitachi Medical Corporation, Kashiwa, Chiba}, 
Masashi TAKIGAWA$^{1}$\thanks{E-mail address: masashi@issp.u-tokyo.ac.jp}, 
Yuta SAIGA$^{1, 2}$, Yoshiya UWATOKO$^{1}$ Hiroshi KAGEYAMA$^{3}$ and Yutaka UEDA$^{1}$}

\inst{$^{1}$Institute for Solid State Physics, The University of Tokyo, Kashiwa, Chiba 277-8581

$^{2}$Department of Physics, Saitama University, Saitama 338-8570

$^{3}$Department of Chemistry, Graduate School of Science, Kyoto University, Kyoto 606-8502}

\recdate{\today}

\abst{We report results of $^{11}$B NMR and susceptibility measurements on the 
quasi 2D frustrated dimer spin system SrCu$_{2}$(BO$_{3}$)$_{2}$ under high 
pressure.  At 2.4~GPa and in a magnetic field of 7~T, NMR lines split with decreasing 
temperature in two steps.  A gradual splitting below $T$=30~K breaking the four-fold 
symmetry of magnetic response is followed by a further sudden splitting below 3.6~K.  
The latter indicates a magnetic phase transition, which is also marked by a kink in the 
susceptibility at 1.44~GPa.  From the magnetic hyperfine shift data, we conclude that 
the low-$T$ phase has a doubled unit cell containing two types of dimers, one in a 
nearly singlet state and the other with a finite magnetization down to $T$=0.}

\kword{SrCu$_{2}$(BO$_{3}$)$_{2}$, Shastry-Sutherland model, high pressure, NMR, phase transition}

\begin{document}

\sloppy
\maketitle

\begin{figure}[b]
 \begin{center}
  \includegraphics[width=80mm]{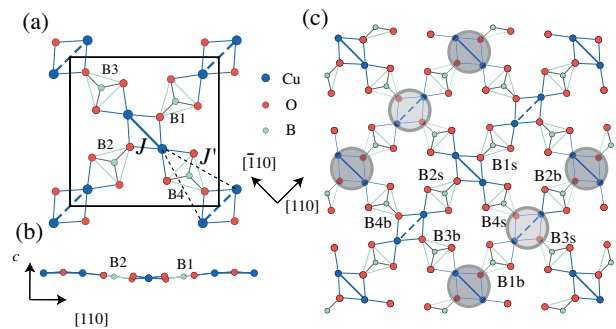}
  \caption{(Color online) The magnetic layer of SrCu$_{2}$(BO$_{3}$)$_{2}$
  viewed along (a) the $c$-direction and (b) the [$\overline{1}$10]-direction.  
  (c) A possible ordered structure in the low-$T$ phase.   Shaded circles represent 
  the gmagnetich dimers.}
  \label{fig:Cryst}
 \end{center}
\end{figure}

%\section{Introduction}
A variety of exotic phenomena has been discovered in the quasi two dimensional 
dimer spin system SrCu$_{2}$(BO$_{3}$)$_{2}$\cite{Kageyama991,Miyahara031}. 
It has an alternating stack of the magnetic CuBO$_{3}$ layers (Figs.~\ref{fig:Cryst}(a) 
and \ref{fig:Cryst}(b)) and the non-magnetic Sr layers\cite{Smith911,Sparta011}. 
The magnetic layer containing orthogonal arrays of spin-1/2 Cu$^{2+}$ dimers 
is a realization of the 2D Shastry-Sutherland model\cite{Shastry811},    
\begin{equation}
H=J\sum_{n.n.}\mathbf{S}_{i} \cdot \mathbf{S}_{j}+J'\sum_{n.n.n.}\mathbf{S}_{i} \cdot \mathbf{S}_{j},
\end{equation}
where $J$ ($J^{\prime}$) is the intradimer (interdimer) Heisenberg exchange interaction. 
The ground state of this model is obvious in two limiting cases: 
the dimer singlet phase for $J^{\prime}/J << 1$ and the N\'{e}el ordered phase for $J^{\prime}/J >> 1$. 
The dimer singlet phase is known to be stable up to $\left( J^{\prime}/J \right)_{c}$=0.68\cite{Miyahara991,Weihong991,Koga001}. 
Various experiments have established that SrCu$_{2}$(BO$_{3}$)$_{2}$ has a dimer singlet ground state
at ambient pressure and zero magnetic field\cite{Kageyama991,Kageyama001,Kodama021} with the 
energy gap of 33~K\cite{Nojiri031,Kakurai021,Gaulin041} and $J^{\prime}/J$=0.60-0.64\cite{Miyahara001,Knetter001}.  

Frustration in the Shastry-Sutherland model strongly suppresses the 
kinetic energy of triplets\cite{Miyahara991}.  Indeed SrCu$_{2}$(BO$_{3}$)$_{2}$ has an 
extremely small width of the triplet dispersion ($\sim$0.2meV\cite{Kakurai021,Gaulin041}).  
Such localized nature of triplets leads to formation of various bound states
of two triplets\cite{Totsuka011,Knetter001} as observed by Raman\cite{Lemmens001}
and neutron~\cite{Aso051} scattering.  It also leads to the magnetization 
plateaus at 1/8, 1/4, and 1/3 of the saturated magnetization
in high magnetic fields\cite{Kageyama011}, where triplets crystalize in commensurate superlattices 
due to mutual repulsion\cite{Kodama022,Miyahara032,Takigawa041,Takigawa061}.  

Since $J^{\prime}/J$ in SrCu$_{2}$(BO$_{3}$)$_{2}$ is close to the critical value,
tuning the exchange parameters, e.g. by applying pressure, might enable us to explore the phase diagram
of the Shastry-Sutherland model, which is still an open issue.  A plaquette singlet phase was proposed to 
exist between the dimer singlet and the N\'{e}el ordered phase\cite{Koga001,Koga002}. 
Alternatively, instability of two-triplet bound states\cite{Momoi001} may lead to a spin \textit{nematic} phase. 
Further variation may arise from the Dzyaloshinski-Moriya interaction beyond the 
Shastry-Sutherland model \cite{Kodama051}.

\begin{figure*}[t]
 \begin{center}
  \includegraphics[width=170mm]{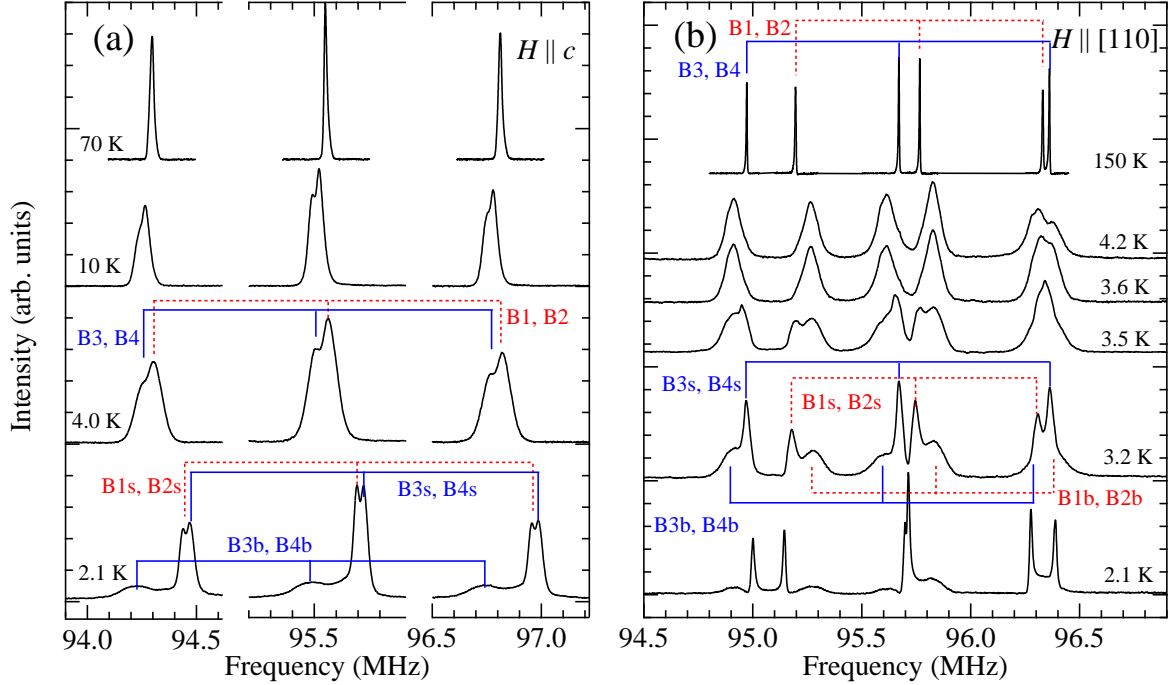}
  \caption{(Color online) Variation of the NMR spectra with temperature at 2.4~GPa for (a) 
$\mathbf{H} \parallel \mathbf{c}$ and (b) $\mathbf{H} \parallel [110]$ at the magnetic field $H$=7.006~T,  
The peaks assigned to B1 and B2 (B3 and B4) are marked by the dashed red (solid blue) lines.}
  \label{fig:Spec}
 \end{center}
\end{figure*}
In spite of such interest, only a few experiments under pressure have been reported to date.  
Magnetic susceptibility data up to $P$=0.7 GPa indicates reduction of the energy gap 
extrapolating to zero near $P$=2.5-3.0~GPa\cite{Kageyama031}.  The X-ray study shows 
a tetragonal to monoclinic structural transition at 4.7~GPa\cite{Loa051}. 
In this letter, we report results of the nuclear magnetic resonance (NMR) experiments on $^{11}$B nuclei 
at $P$=2.4~GPa and the susceptibility measurements up to $P$=1.44~GPa.  
Our data provide evidence for a magnetic phase transition below 4~K into an 
ordered phase with two distinct types of dimers.  
  
A single crystal of SrCu$_{2}$(BO$_{3}$)$_{2}$ prepared by the traveling-solvent-floating-zone 
method\cite{Kageyama992} was cut into a thin plate (2.0$\times$2.8$\times$0.3mm$^{3}$) for NMR 
measurements to reduce distribution of demagnetizing field. It was placed in a piston-cylinder-type 
pressure cell made of NiCrAl and BeCu alloys filled with 1:1 mixture of N-pentane and isoamyl-alcohol.  
The pressure was calibrated against the load applied at room temperature by separate 
measurements of the superconducting transition temperature of Sn metal.  
The pressure cell was mounted on the NMR probe with a double-axis-goniometer to enable  
arbitrary alignment of the crystal in magnetic fields.  The $^{11}$B NMR spectra were obtained by Fourier 
transforming the spin-echo signal.  The demagnetizing field was corrected by comparing the NMR 
frequencies at ambient pressure to the published data obtained on a nearly spherical 
crystal\cite{Kodama021,Kodama051}.  The susceptibility was measured on a different crystal 
with a SQUID magnetometer (Quantum Design, MPMS) equipped with a BeCu pressure 
cell\cite{Uwatoko051} using Daphne Oil 7373 as the pressure transmitting fluid.  

Figure~\ref{fig:Spec} shows the NMR spectra under pressure ($P$=2.4~GPa) at various temperatures ($T$) 
in the field $H$=7.006~T applied along (a) the $c$- and (b) the [110]-directions.  
As $^{11}$B nuclei have spin 3/2, frequencies of the quadrupole-split three NMR lines are given as\cite{Slichter}, 
\begin{equation}
\nu_{m \leftrightarrow m-1} = \left( 1+K \right) \gamma H + \left( m-1/2 \right) \nu_{Q} + \delta\nu^{(2)}_{m} , 
\label{resonance}
\end{equation}
$m$=3/2, 1/2, or -1/2.  Here $\gamma$=13.66~MHz/T is the nuclear gyromagnetic ratio and $K$ is 
the magnetic hyperfine shift caused by the coupling between nuclei and magnetization on neighboring
Cu sites.  The second term is the first order quadrupole shift with $\nu_{Q}$ proportional 
to the electric field gradient (EFG) along the magnetic field direction.  
This term vanishes for the central line ($m$=1/2). The third term, the second 
order quadrupole shift, is identical for the two satellite lines ($m$=3/2 and -1/2).    

SrCu$_{2}$(BO$_{3}$)$_{2}$ has tetragonal structure with the space group $I\overline{4}2m$
at ambient pressure and temperatures below 395K \cite{Smith911,Sparta011}.  The Cu and B atoms 
both occupy a unique 8$i$ site located on the (110) or ($\overline{1}$10) mirror plane (Fig.~\ref{fig:Cryst}(a)).  
A unit cell contains two magnetic CuBO$_{3}$ layers related by the translation $t$(1/2, 1/2, 1/2).  
The four B atoms in a unit cell per layer, B1 - B4 in Fig.~\ref{fig:Cryst}(a), give distinct NMR 
frequencies for general field directions.  The number of NMR lines is reduced for symmetric 
directions.  When the field $\mathbf{H}$ is in the $(\overline{1}10)$ mirror plane containing  
the $c$- and the [110]-directions, B3 and B4 sites are equivalent but B1 and B2 are not 
due to buckling of CuBO$_{3}$ layers (Fig.~\ref{fig:Cryst}(b))\cite{Kodama051}.  
Then B1, B2 and (B3, B4) give three sets of quadrupole split three lines.  For $\mathbf{H} \parallel [110]$, 
B1 and B2 also become equivalent resulting in two sets of lines.  For $\mathbf{H} \parallel \mathbf{c}$, 
all four sites are equivalent.  The NMR spectra at ambient pressure are indeed consistent with these 
predictions at all temperatures\cite{Kodama021,Kodama051}.  

\begin{figure*}[t]
 \begin{center}
 \includegraphics[width=170mm]{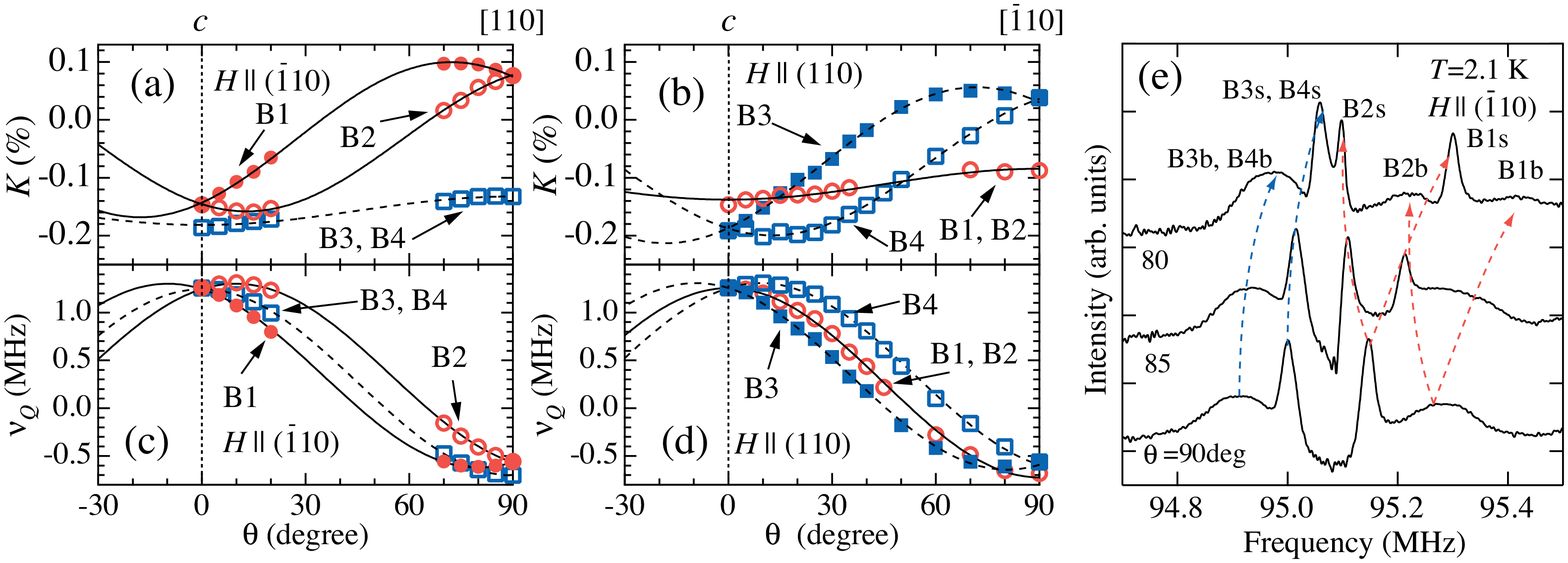}
  \caption{(Color online) (a) - (d): Angle dependences of $K$ and $\nu_{Q}$ at $T$=10~K.  
The lines show the fitting explained in the text.  (e): Angular variation of the NMR spectum at $T$=2.1~K 
with the field in the ($\overline{1}$10)-plane.  Only the low frequency satellite lines 
($m$=-1/2) are displayed for clarity. Intensity is plotted in a logarithmic scale to make the 
broad lines clearly visible }
  \label{fig:rot10K}
 \end{center}
\end{figure*}

At $P$=2.4~GPa, only one set of NMR lines is observed at high temperatures for 
$\mathbf{H} \parallel \mathbf{c}$ (Fig.~\ref{fig:Spec}(a)), consistent with the crystal symmetry at 
ambient pressure.  Upon cooling below 30~K, however, each line begins to split gradually 
and gets broadened.  All three quadrupole split lines show clear double peak structure at 10~K and 
4~K (Fig.~\ref{fig:Spec}(a)).  In order to make site assignment for the split peaks, we examined 
variation of the spectra with the field rotated in the $(\overline{1}10)$-plane at 10~K.  We found 
that one of the split peaks further splits into two lines, while the other peak remains unsplit.  
The unsplit peak was then assigned to (B3, B4) and each of the split lines to 
B1 or B2.  We repeated the measurements for the field rotated in the (110)-plane.  The lines 
assigned to B3 and B4 split but lines from B1 and B2 do not, as expected.  

The value of $\nu_{Q}$ determined from the spacing between the two satellite lines is plotted 
against the angle $\theta$ between $\mathbf{H}$ and the $c$-direction in Fig.~\ref{fig:rot10K}(c) 
for $\mathbf{H}\parallel(\overline{1}10)$ and in Fig.~\ref{fig:rot10K}(d) for $\mathbf{H}\parallel(110)$.
The $\theta$-dependence of $K$ is then determined from the average 
frequency of the two satellite lines after subtracting $\delta\nu^{(2)}_{m}$ calculated from the 
$\nu_{Q} (\theta)$ data\cite{Stauss} as shown in Figs.~\ref{fig:rot10K}(a) and \ref{fig:rot10K}(b). 

The distinction between (B1, B2) and (B3, B4) revealed by the line 
splitting for $\mathbf{H} \parallel \mathbf{c}$ must be ascribed to the loss of four fold symmetry 
($\overline{4}$) around the $c$-direction.  This symmetry requires that $\nu_{Q}(\theta)$ and 
$K(\theta)$ at  B1 and B2 (B3 and B4) for $\mathbf{H} \parallel (\overline{1}10)$ be identical to those at  
B3 and B4 (B1 and B2) for $\mathbf{H} \parallel (110)$.  The data in Fig.~\ref{fig:rot10K} show 
that this condition is grossly violated for the magnetic shift $K$ but not for the quadrupole 
coupling $\nu_{Q}$.  This strongly suggests that the symmetry change is primarily due to magnetic 
origin since any structural change should be better sensed by $\nu_{Q}$.  Thus we conclude that the two 
sublattices of orthogonal Cu dimers shown by the solid and dashed lines in Fig.~\ref{fig:Cryst}(a) 
become inequivalent with different magnetizations.  We expect though this may accompany a slight 
structural change.  Detailed structural analysis is left for future studies. 

The $K(\theta)$ and $\nu_{Q}(\theta)$ data can be fit to the standard formula for anisotropic shifts,
$u+v\cos^{2}(\theta-\alpha)$ with $u$, $v$, and $\alpha$ being the fitting parameters\cite{Slichter}, 
as shown by the lines in Figs.~\ref{fig:rot10K}(a) - \ref{fig:rot10K}(d).  
We found that $K(\theta)$ and $\nu_{Q}(\theta)$ 
at B1 for $\mathbf{H}\parallel(\overline{1}10)$ (at B3 for $\mathbf{H}\parallel(110)$) are identical to 
$K(-\theta)$ and $\nu_{Q}(-\theta)$ at B2 (at B4).  Thus the mirror symmetries are preserved.  
The loss of $\overline{4}$ changes the space group from $I\overline{4}2m$ to 
orthorhombic $Fmm2$.  Our data indicate that the entire crystal forms a single domain. 
 
We now discuss the NMR spectra in Fig.~\ref{fig:Spec} at lower temperatures.  
For $\mathbf{H} \parallel [110]$, (B1, B2) and (B3, B4) give distinct lines at all 
temperatures. No line splitting is observed down to 3.6~K.  At 3.5~K, however, all lines 
develop clear two peak structure.  With further decreasing temperature, these two peaks 
change into one sharp and one broad lines with nearly equal intensity denoted as B$n$s 
and B$n$b ($n$=1 - 4) in Fig.~\ref{fig:Spec}(b).  Figure \ref{fig:rot10K}(e) shows the 
variation of the low frequency satellite lines ($m$=-1/2) when the field is rotated 
from [110] toward the $c$-direction at 2.1~K.  Both the sharp and the broad lines 
from (B1, B2) split in a similar manner as observed at higher temperatures.  Therefore, each 
of B1 and B2 must be divided into two sites below 3.6~K, (B1s, B1b) and (B2s, B2b), yielding  
eight inequivalent B sites for general field directions.  A Similar spectrum with sharp 
and broad lines is observed also for $\mathbf{H} \parallel \mathbf{c}$ at 
2.1~K (Fig.~\ref{fig:Spec}(a)), although there is only one set of broad lines. 
We found that this belongs to (B3, B4), while the broad lines from 
(B1, B2) overlap with the sharp lines, by extending the measurements shown in 
Fig.~\ref{fig:rot10K}(e) to smaller values of $\theta$.  

\begin{figure*}[t]
\begin{center}
 \includegraphics[width=170mm]{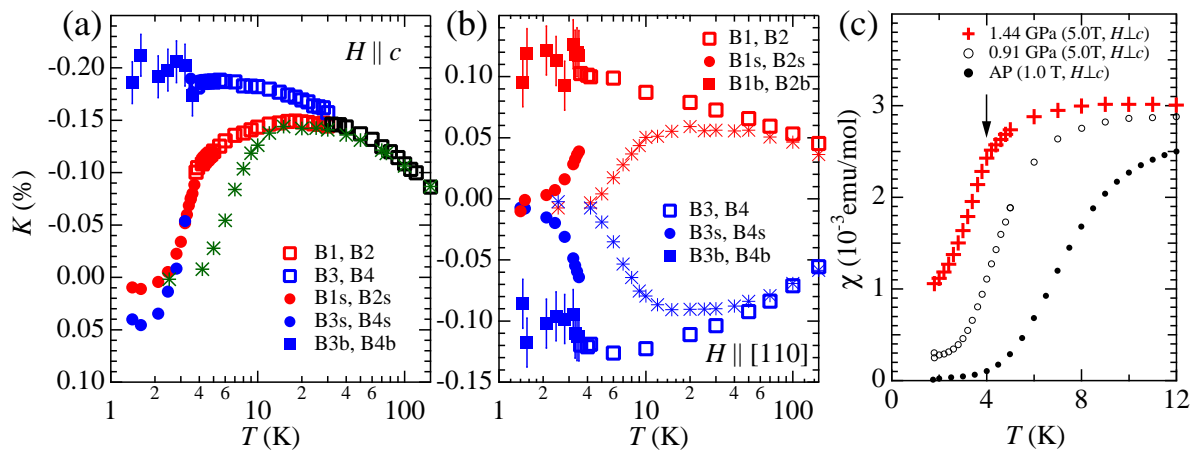}
 \caption{(Color online) $T$- dependences of the shifts at 2.4~GPa for (a) $\mathbf{H} \parallel \mathbf{c}$ 
and (b) $\mathbf{H} \parallel [110]$ compared with the data at ambient pressure shown by crosses.
(c): $T$-dependence of the susceptibility at ambient pressure (AP), 0.91~GPa and 1.44~GPa.}
\label{fig:shift_chi}
\end{center}
 \end{figure*}
Figure~\ref{fig:shift_chi} shows the $T$-dependence of the shifts at various sites for (a)
$\mathbf{H} \parallel \mathbf{c}$ and (b) $\mathbf{H} \parallel$ [110] compared with the data at ambient 
pressure\cite{Kodama021,Kodama051}.  Above 40~K, the results at 2.4~GPa are nearly 
unchanged from ambient pressure.  Line splitting appears for $\mathbf{H} \parallel \mathbf{c}$ below 30~K 
as mentioned above.  In spite of a clear change of symmetry, the splitting develops gradually without 
sign of a phase transition.  In contrast, the second splitting at 3.6~K occurs suddenly and clearly marks 
a phase transition.  The shifts for the sharp lines approach near zero as $T \rightarrow 0$, pointing to a
singlet ground state.  We can indeed fit the data to an activation law, $\alpha + \beta \exp (-\Delta/T)$, 
yielding $\Delta$=11-15~K.  These values are much smaller than the gap at ambient 
pressure (24~K) at the same field of 7~T.  The shifts for the broad lines, on the other hand, maintain large 
values down to the lowest temperature, pointing to a magnetic state without an excitation gap.   

These results indicate coexistence of  gmagnetich and  gnon-magnetich Cu dimers in the low-$T$ phase. 
The sharp (broad) lines should come from those B sites which couple dominantly to the non-magnetic 
(magnetic) Cu dimers.  Preliminary results at different fields show that both the hyperfine field 
($K$ multiplied by $H$) and the line width for the broad lines are approximately proportional to the field, indicating 
no spontaneous moment at zero-field.  The increased number of NMR lines indicates doubling of the 
primitive unit cell in the low-$T$ phase.  It is most likely that each of the Cu dimer sublattices 
develop spatial order of magnetic and non-magnetic dimers, forming either a superstructure in the $ab$-plane  
(see Fig.~\ref{fig:Cryst}(c) for an example) or alternating magnetic and non-magnetic layers along
the $c$-direction. 
  
The susceptibility data are presented in Fig.~\ref{fig:shift_chi}(c).  While no anomaly is observed at 
ambient pressure and at 0.91~GPa, the data at 1.44~GPa show a clear kink at 4.0~K, providing 
further evidence for bulk nature of the phase transition.  The slightly different transition temperature 
is presumably due to the difference in magnetic field. Note that the susceptibility approaches a finite 
value as $T \rightarrow 0$ consistent with the coexistence of two types of Cu sites.  
 
What is the order parameter describing the low-$T$ phase ? Since the magnetic dimers 
appears to have no spontaneous moment at zero-field but larger susceptibility than the 
non-magnetic dimers, a natural candidate would be the staggered component of the two-spin 
correlation $\langle \mathbf{S}_{1} \cdot \mathbf{S}_{2} \rangle$ within a dimer.  This is invariant 
under time-reversal and considered a bond-nematic order parameter.  Recently, a bond-nematic 
order has been proposed for frustrated spin systems on a square lattice as a result of Bose 
condensation of two-magnon bound states\cite{Shannon061}.  Whether such a scenario is 
relevant for SrCu$_2$(BO$_3$)$_2$ is an interesting issue.  

To conclude, we have demonstrated that SrCu$_2$(BO$_3$)$_2$ under pressure exhibits 
symmetry lowering in two steps. A gradual loss of four-fold symmetry near 30~K is 
followed by a clear phase transition below 4~K. We propose that the low-$T$ phase has spatial 
order of two types of dimers: one is nearly in a singlet state while the other 
has a finite susceptibility down to  $T$=0. 

%acknowledgment
We thank S. Miyahara, F. Mila, T. Momoi and M. Oshikawa for stimulating discussions and T. Matsumoto for 
help in designing the pressure cell. This work was supported by Grant-in-Aid for COE Research 
(No. 12CE2004) from the MEXT Japan. 
 
 %References
%\section*{References}

\end{document}